# Dislocation Arrays Hosting Topological Bound States in Epitaxial Tetragonal FeSe Films

*Huimin Zhang*[¶,§], *Basu Oli*[¶], *Zhengfei Wang*[†*], *Michael Weinert*[‡], *and Lian Li*[¶*]

¶Department of Physics and Astronomy, West Virginia University, Morgantown, WV 26506, USA

§Key Laboratory of Materials Modification by Laser, Ion and Electron Beams (Dalian University of Technology), Ministry of Education, Dalian 116024, China

†Hefei National Research Center for Physical Sciences at the Microscale, CAS Key Laboratory of Strongly-Coupled Quantum Matter Physics, Department of Physics, University of Science and Technology of China, Hefei, Anhui 230026, China

‡Department of Physics, University of Wisconsin, Milwaukee, WI 53211, USA

*Corresponding author: lian.li@mail.wvu.edu, zfwang15@ustc.edu.cn

**ABSTRACT**

Topological defects in real space can host robust, localized quantum states for quantum information processing and sensing, yet deterministic control over their formation in crystalline materials remain challenging. Here, we show that symmetry-mismatched epitaxy of tetragonal FeSe films on $SrTiO_3$(111) naturally generates networks of screw dislocations and tilt grain boundaries composed of periodic arrays of edge dislocations. Low-temperature scanning tunneling microscopy/spectroscopy measurements reveal localized electronic states pinned to these dislocation cores. Combining with first-principles calculations, we identify these states as topological bound states arising from the interplay between the real-space topological invariant of the defect, the Burgers vector, and the nontrivial electronic topology of FeSe. Our work demonstrates symmetry mismatch epitaxy as a scalable route toward template topological defect networks and their associated quantum states in solid-state materials.

## INTRODUCTION

Defects in crystalline materials, once regarded as detrimental to their functionality, are now central to quantum information processing and sensing because they can host robust, localized states with long spin coherence, optical addressability, and stability under ambient conditions[1,2]. Notable examples include point defects in wide-bandgap semiconductors, such as the nitrogen-vacancy center in diamond[3,4], similar point defects in silicon carbides[5], as well as rare-earth dopants in complex oxides[6,7]. The robustness arises primarily from local properties, including strong chemical bonding, large host bandgaps, and defect-specific point-group symmetries that isolate their electronic levels from the bulk, which in turn suppresses decoherence[2–7].

Beyond the point defect platforms, the advent of topology in materials science has introduced a fundamentally different mechanism for robustness rooted in global topological invariants defined in both real and momentum space[8,9]. The topological classification of electronic band structures has revealed a novel class of quantum materials that host boundary states protected by bulk topological indices and therefore remain stable against all local perturbations that do not close the bulk gap[10]. Similarly, topological lattice defects, such as dislocations, carry quantized topological charges (e.g., Burgers vectors) that local, continuous deformations of the lattice structure cannot remove[11]. When such defects are embedded in a material with nontrivial band topology, the coupling between their real-space topological charges and the bulk topological band structures can give rise to a range of intriguing phenomena, most notably robust bound states characterized by fractional quantum numbers[12–14].

Furthermore, introducing extended defects in crystals, such as the tilt grain boundary (GB) illustrated in Figure 1a, can generate periodic arrays of these localized states[15]. In the small-angle limit, the GB can be described as a periodic array of edge dislocations, characterized by the misorientation angle $\theta$

$$\tan\theta = b/d \qquad (1)$$

where $b$ is the Burgers vector, and $d$ is the spacing between dislocations.

A $\pi$-phase shift is expected at an edge dislocation with a Burgers vector $b = (a, 0)$, leading to a sign change in the position-dependent Dirac mass $\boldsymbol{m(\theta)}$ and the formation of a mass domain-wall at the dislocation (Figure 1b-c). Following the Jackiw-Rebbi model[16], such a domain wall supports a localized gapless helical mode (red curve in Figure 1c), rendering the GB an intrinsic 1D topological channel that hosts periodic topological defect-bound states within the bulk.

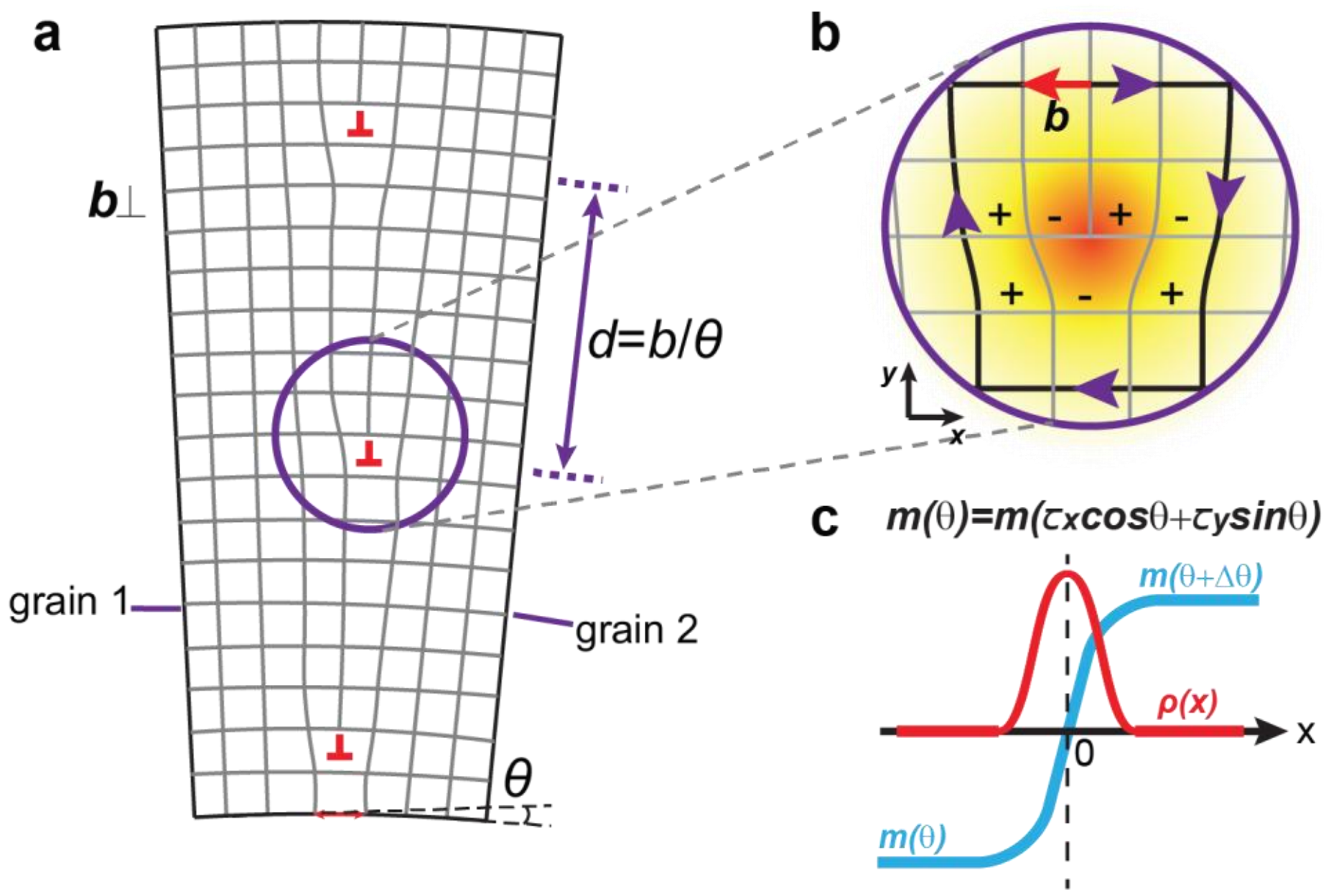

**Figure 1. Periodic dislocation-bound states in a low-angle grain boundary. (a) Schematic illustration of the dislocation model of low-angle GB characterized by a misorientation angle $\theta$, modeled as a periodic array of edge dislocations. (b) Zoom-in view of an edge dislocation marked by the purple circle in (a). A local electronic mode emerges at the dislocation core due to the π phase difference across the defect. (c) The mass $m(\theta)$ and density of states $\rho(x)$ across the edge dislocation shown in (b). The π phase shift leads to a sign-change in the Dirac mass $m(\theta)$ across the dislocation, giving rise to a bound state represented by the peak in the local density of states $\rho(x)$ (red curve).**

To date, the realization of these localized topological states has been primarily in photonic and metamaterial systems, including modes with fractional charge bound to topological disclinations and protected by the lattice rotational symmetry[14,17–25]. Recent theoretical studies have further examined the influence of structural defects, such as vacancies, on the topological insulator phases[26], and the quantized ON/OFF conductance of helical/chiral dislocation states has been proposed for topological field-effect transistors[27]. However, solid-state realizations, especially in correlated and superconducting materials, remain limited[28]. The primary obstacle has been the lack of material platforms that simultaneously exhibit nontrivial bulk topology and allow controlled formation of real-space topological defects.

To address this challenge, we use molecular beam epitaxy (MBE) to grow tetragonal FeSe(001) on the six-fold symmetrical $SrTiO_3$(111) substrate, which promotes the formation of growth spirals and low-angle GBs composed of periodic edge dislocations. Tetragonal FeSe hosts a rich set of correlated and topological electronic phases[29]: it undergoes a nematic transition near 90 K without magnetic order, and its superconductivity is highly tunable from 8–9 K in bulk to >70 K in monolayer FeSe/$SrTiO_3$(001). Moreover, FeSe/$SrTiO_3$(001) is a 2D topological insulator[30–32], while multilayer FeSe further exhibits Luttinger liquid behavior on its edge channel[33]. These properties make FeSe an excellent platform for probing the interplay between crystal topological defects and electronic band topology. We show that spiral growth of tetragonal FeSe(001) on the $SrTiO_3$(111) substrate produces screw dislocations and low-angle GBs with regular edge-dislocation arrays. Low-temperature scanning tunneling microscopy/spectroscopy directly reveals localized electronic states bound to dislocation cores. First-principles calculations confirm that these states originate from the nontrivial bulk topology of tetragonal FeSe. Together, these results establish topological dislocation states in FeSe. Beyond the FeSe/$SrTiO_3$ system, our work demonstrates a general heteroepitaxial strategy in which mismatch in rotational symmetry between the film and substrate enables the controlled formation of dislocations and GBs networks with tunable densities, providing a template for engineering topology-protected bound states. The approach is compatibility with wafer scale thin-film growth and readily applicable to van der Waals and oxide heterostructures, opening a scalable route for predictive design of quantum defects for next-generation quantum technologies.

## RESULTS AND DISCUSSION

Tetragonal FeSe (*P4/nmm*) is the structurally simplest iron-based superconductor composed of stacked layers of edge-sharing $FeSe_4$ tetrahedra with strong in-plane covalent bonding and weak interlayer van der Waals interactions. This anisotropic bonding enhances structural compliance and tolerance to lattice mismatch, enabling epitaxial FeSe films to accommodate strain through the formation of spiral morphologies and dislocation networks when grown on symmetry-mismatched substrates[34–36]. On the six-fold symmetric $SrTiO_3$(111) substrate (Figure 2a and Figure S1), the four-fold tetragonal FeSe(001) lattice can align such that its principal axes coincide with those of the substrate[37,38], yielding three equivalent rotational domains, each relative to the others by 120°. As these domains expand laterally during growth, they inevitably coalesce to form two types of topological defects: 1) GBs when merging occurs on terraces and 2) growth spirals when coalescence takes at step edges.

STM imaging shows that the GBs appear as arrays of characteristic dots (one marked by the arrow in Figure 2c and Figure S2). Atomic-resolution imaging reveals these features as edge dislocations, as displayed in Figures 2e-f, separating the surface into two domains where the in-plane Se lattice is rotated by a finite angle, consistent with the dislocation model for low-angle GB (Eq. (1)). The rotation angle can be extracted from two lines drawn across the boundary (Figure 2e), or from pattern generated from the fast Fourier transform (FFT) of the STM image (Figure 2f). Away from GB, the (1 × 1) tetragonal lattice is clearly resolved with a spacing of 3.9 Å. By tracing the atomic lattice around a dislocation core, as shown in Figure 2f, an in-plane Burgers vector of [$a_{Se}$, 0] is determined. As the rotation angle $\theta$ increases, the dislocation spacing decreases (Figure 2e and Figure S3), as expected from the low-angle GB model (Eq. 1).

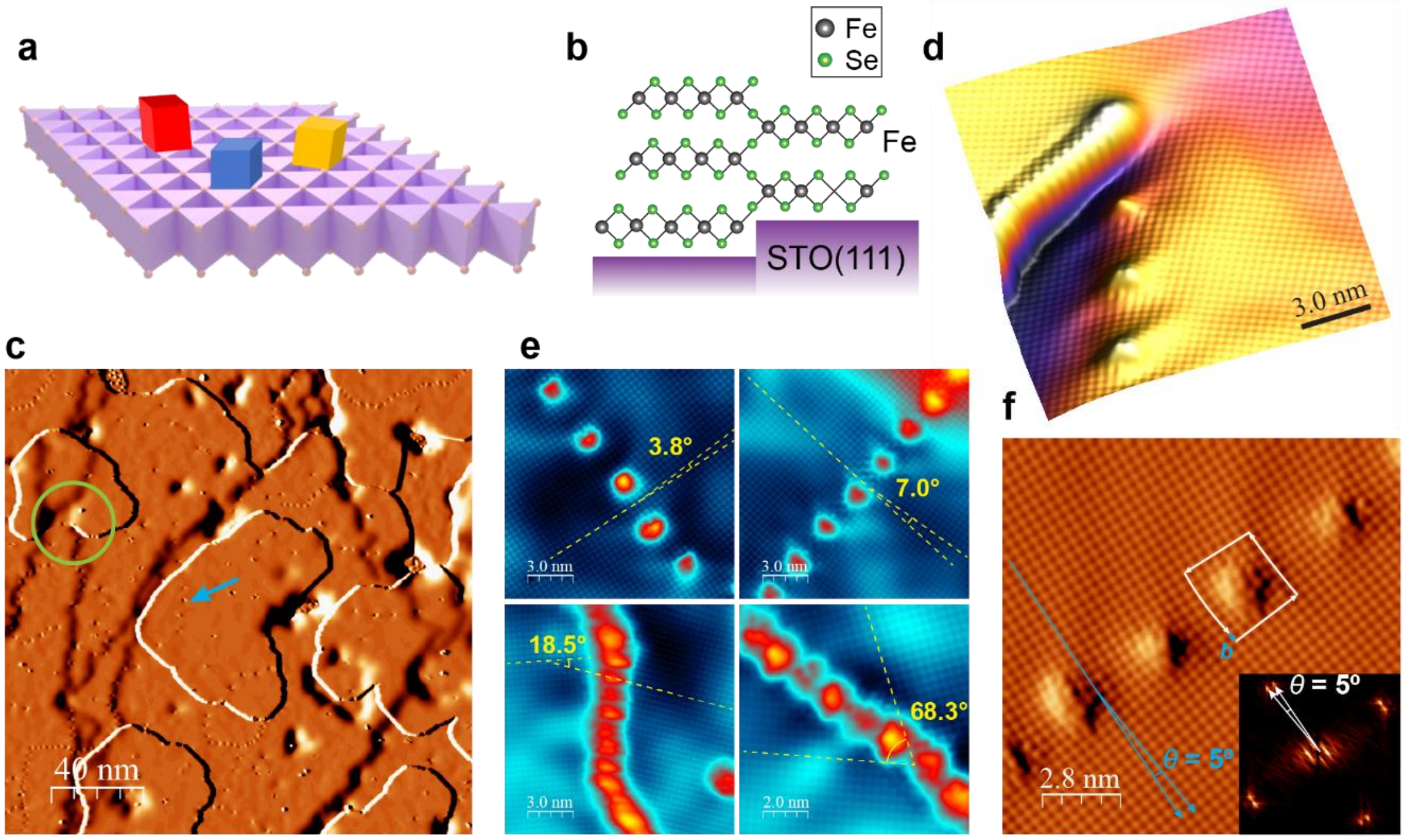

**Figure 2. Spiral growth of FeSe films on the $SrTiO_3$(111) substrate. (a) Schematic diagram of the tetragonal FeSe lattice epitaxially aligned on the $SrTiO_3$(111) substrate. The red, blue, and yellow cubes represent the three rotational orientations of the FeSe lattice on STO(111).**

**(b) Schematic illustration of screw dislocation formation on STO(111) substrate. (c) Differential-mode STM image, highlighting the presence of screw dislocations (one circled) and low-angle GBs, one marked by the arrow. The dotted, periodically spaced contrast variations correspond to edge dislocations. Setpoint: V = 1.0 V, I = 100 pA. (d) 3D representation of a screw dislocation, showing a Burgers vector out-of-plane. (e) STM images of four low-angle GBs with the rotation angle θ across grain boundaries labeled. Setpoint: V = 150 mV, I = 0.1 nA (left top), V = 30 mV, I = 0.3 nA (right top), V = 40 mV, I = 2.0 nA (left down), V = 50 mV, I = 1.0 nA (right down). (f) Atomic-resolution STM image of low-angle GB in differential mode. Setpoint: V = 100 mV, I = 100 pA. The burgers vector b = [a, 0, 0] is indicated by the cyan arrow. Inset: FFT pattern of the atomic resolution image.**

A second type of topological defect emerges when neighboring FeSe domains coalesce at step edges of the $SrTiO_3$(111) substrate. As illustrated in Figure 2b, this process results in a vertical lattice mismatch in which the top Se layer of the lower terrace bonds with the bottom layer of the upper terrace, generating a screw dislocation with a Burgers vector oriented along the *c*-axis. This dislocation subsequently drives spiral growth in which atomic layers wind continuously around the core, forming a chiral step-terrace structure readily resolved by STM (one example is circled in Figure 2c and Figure S1f). Atomic-resolution STM imaging (Figure 2d) confirms a step height of ~0.4 nm, consistent with a single FeSe trilayer. Away from the dislocation core, the lattice is well-ordered with the (1 × 1) square lattice of 3.9 Å, characteristic of tetragonal FeSe(001). Interestingly, the density of spirals decreases as the film grows thicker, owing to the merging of adjacent spirals, a behavior reminiscent of spiral growth observed in $Bi_2Se_3$ films on graphene/SiC(0001)[34].

Together, the spiral growth and the formation of low-angle GBs yield three recurrent structural motifs: (i) single-layer steps separating atomically flat terraces, (ii) screw dislocations, and (iii) periodic edge dislocation arrays. Using systematic differential conductance (dI/dV) measurements, we investigate regions near step edges as well as the cores of screw and edge dislocations, enabling direct visualization of spatially localized bound states in FeSe(001) films (Figures (a-c)).

First, a single spiral yields three successive layers separated by single-layer steps, as shown in Figure 3a, The dI/dV spectra taken on the terraces exhibit two key features: (1) a pronounced peak below the Fermi level (black arrows), appearing between −150 and −200 meV, attributed to the Fe $3d_z^2$ bands at the $\Gamma$ point[39,40]; and (2) a significant suppression of the density of states (DOS) near the Fermi level, producing an apparent gap of ~Δ = 52 meV (Figure 3b and Figure S4) that is likely associated with nematic order commonly observed in multilayer FeSe(001) films[34]. As discussed below, this gap facilitates the detection of topological edge and dislocation-bound states by making these localized modes more visible.

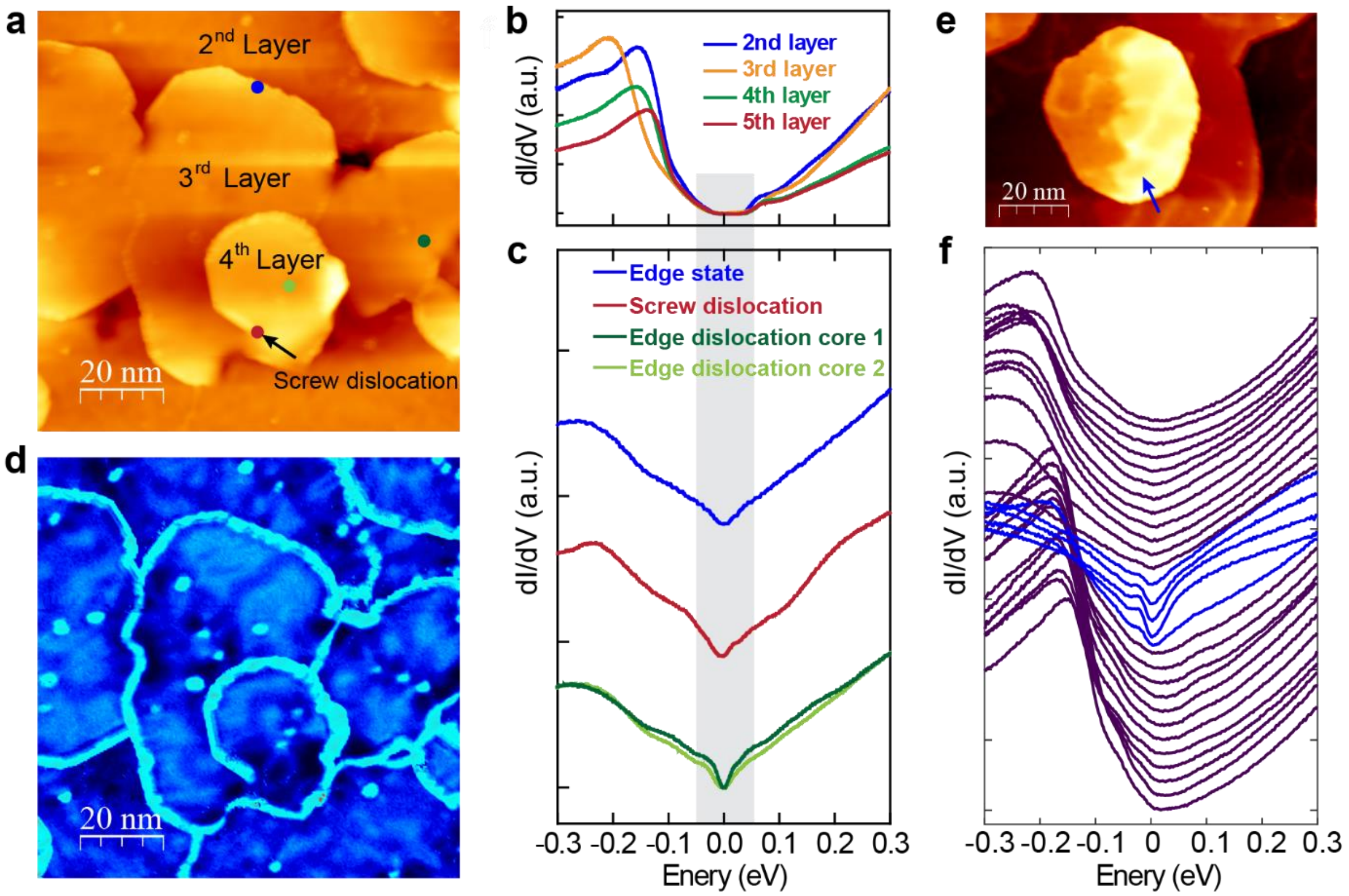

**Figure 3. Edge and dislocation modes in FeSe/SrTiO$_3$(111). (a) Topographic STM image showing the grain boundaries and screw dislocations in FeSe/STO(111) films. (b) Representative dI/dV spectra taken at different layers of FeSe/STO(111) films. (c) dI/dV spectra measured at the film edge, screw dislocation core, and edge dislocation cores, with the corresponding sites marked by color-coded dots. (d) dI/dV map corresponding to the topographic image in (a). (e) Topographic STM image showing the edge of a FeSe island. Setpoint: *V* = 1.0 V, *I* = 30 pA. (f) Line dI/dV spectra taken along the blue arrow in (e).**

Near step edges, the electronic structure shows pronounced changes (Figure 3c), including suppression of the -200 meV peak and a transition from the gap to V-shaped spectra near $E_F$, indicative of metallic edge channels[33]. The spatial distribution of these edge states was further examined by $dI/dV$ conductance maps $g(\mathbf{r}, E)$, acquired simultaneously with the topography image. As shown in Figure 3d, the $g(\mathbf{r}$, 50 meV) reveals strong conductance enhancement along the step edges of the third and fourth layers. To quantify their spatial extent, $dI/dV$ spectra were taken along line cuts normal to the edge (Figure 3e), spanning the second to third FeSe layers. These measurements indicate that the edge states are confined within ~2 nm of the step edge (Figure 3f).

This behavior is consistent with previous studies identifying single-layer FeSe as a 2D TI, in which symmetry-protected edge states arise from a combined invariant of time-reversal symmetry and lattice translation[31]. In multilayer FeSe, stacking these 2D topological layers along the out-of-plane *c* axis realizes a weak topological insulating phase, with step edges acting as effective side-surface terminations that host the corresponding weak-TI boundary modes[41]. In addition, the observed V-shape spectra reflect strong electron-electron interactions consistent with Tomonaga–Luttinger Liquid behavior[33,42,43].

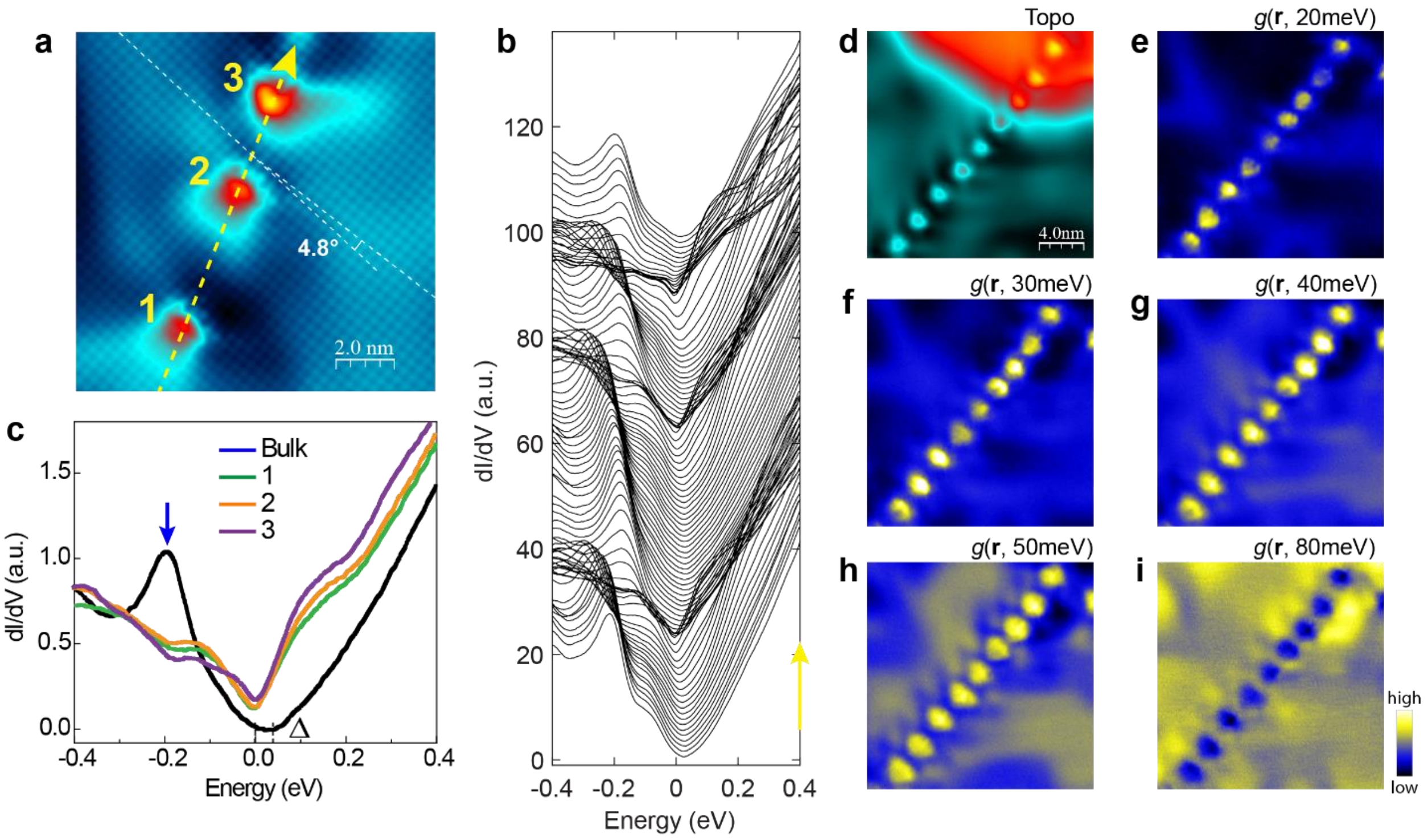

**Figure 4. Spatial dependence of localized states at edge-dislocation cores in FeSe/STO(111) films. (a) Topographic STM image showing three edge dislocations in a FeSe/STO(111) film. Setpoint: *V* = 1.0 V, *I* = 1.0 nA. (b) Line dI/dV spectra acquired along the dashed yellow**

**arrow in (a), illustrating the spatial evolution of the electronic states across the dislocations. (c) Comparison of dI/dV spectra taken at three edge dislocation cores and in the bulk (away from the dislocation). The characteristic dI/dV spectrum of the bulk FeSe exhibits a pronounced peak around -0.2 eV and a gap, Δ, within the energy range [0, 80] meV. In contrast, the spectra at the dislocation cores show a universal V-shaped feature, indicative of localized electronic states. (d-i) Topographic image and corresponding dI/dV maps revealing local modes emerging at the edge dislocation cores. Setpoint: $V$ = 0.1 V, $I$ = 0.5 nA, $V_{mod}$ = 2.0 mV.**

Second, the dI/dV spectra acquired at the core of edge dislocation, labeled 1 and 2 in Figure 3c, exhibit two similar features: 1) a suppression of the spectral peak between -150 and -200 mV, and 2) a partial filling of the bulk gap with V-shaped states centered at the Fermi level. To further quantify their spatial distribution, $dI/dV$ conductance maps $g(\mathbf{r}, E)$, and spatially resolved $dI/dV$ spectra were acquired for a 4.8° GB (Figure 4a). The $dI/dV$ spectra were taken at the core of the three adjacent edge dislocations (Figure 4b), which consistently exhibit the V-shaped states at $E_F$ and the reduced spectral peak at ~-200 meV. $dI/dV$ spectra line scans, taken along the path indicated by the yellow dotted arrow in Figure 4a, demonstrate that the Fe $3d_z^2$ peak intensity diminishes toward the dislocation core, accompanied by the gradual development of a V-shaped spectrum that becomes most pronounced at the core (Figure 4c). This pattern repeats at the adjacent dislocations, confirming the reproducibility and localization of the dislocation states. These features are consistently observed across FeSe multilayers with similar low-angle GBs. Moreover, the differential conductance maps $g(\mathbf{r}, E)$ at selected energies (Figures 4d–i) reveal enhancement of the dislocation cores, further corroborating the existence of localized dislocation states. Note that the nematic domain walls appear as stripe-like features with strong contrast, particularly at 0 meV and –10 meV (Figure S5), consistent with previous report[44]. In contrast, the dislocation-bound states manifest as bright, dot-like features corresponding to an enhanced local density of states confined to the dislocation cores within the energy window of 0 to 60 meV above the Fermi level. Outside this energy range, the dislocation cores exhibit suppressed density of states. The clear energy separation between nematic domain walls and dislocation states, together with the absence of spatial pinning, indicates that the dislocation states are not strongly correlated with the nematic domain walls.

We further examined dislocation associated with GBs of different misorientation angles. As shown in Figure S6, spatially resolved dI/dV spectra taken across both the small-angle ($\theta$=5.2°) and larger-angle ($\theta$=68.3°) GBs are consistently V-shaped. The corresponding dI/dV maps acquired at +50 meV and +100 meV indicates localized dislocation bound states along the GB, while a suppressed density of states is observed at other energies. Importantly, no measurable energy splitting or dispersive bands are observed within the energy resolution and dislocation spacing range investigated, indicating that the neighboring dislocation states remain weakly coupled."

Third, similar dI/dV spectroscopic signatures are observed near screw dislocations (Figure 3c and Figure S7), which exhibit similar V-shaped near $E_F$ and a suppressed negative-bias peak. These shared characteristics across distinct dislocation types reveal a common mechanism by which the real-space topological defects invert the sign of the Dirac mass across the dislocation core, enabling the formation of topological domain-wall bound states in FeSe, as discussed below.

To elucidate the nature of dislocation bound states observed by STM, we performed first-principles density-functional theory (DFT) calculations for a FeSe monolayer to obtain tight-binding parameters. We construct the lattice configuration of the edge dislocation using the established cut-and-glue technique[45] by removing a stripe of FeSe unit cells and gluing the two parts together. Each boundary of the FeSe squares carries a pair of quantum spin Hall edge states spreading in opposite directions, as illustrated by two pairs of blue and red arrows in Figure 5a. When the two boundaries are decoupled (Figure 5a), the edge states exhibit massless Dirac states (Figure 5b). In contrast, for the dislocation configuration (Figure 5c), the edge states from the two FeSe squares interact with each other, resulting in gapping of the Dirac state, i.e., the introduction of a mass, $m(\theta)$ that varies along the boundaries (Figure 5d); the mass angle $\theta$ varies continuously except with a jump of $\pi$ at the center of the dislocation. The sign-change of the mass term $m(\theta)$ forms a 1D domain wall and generates a 0D topological state (Figure 5d).

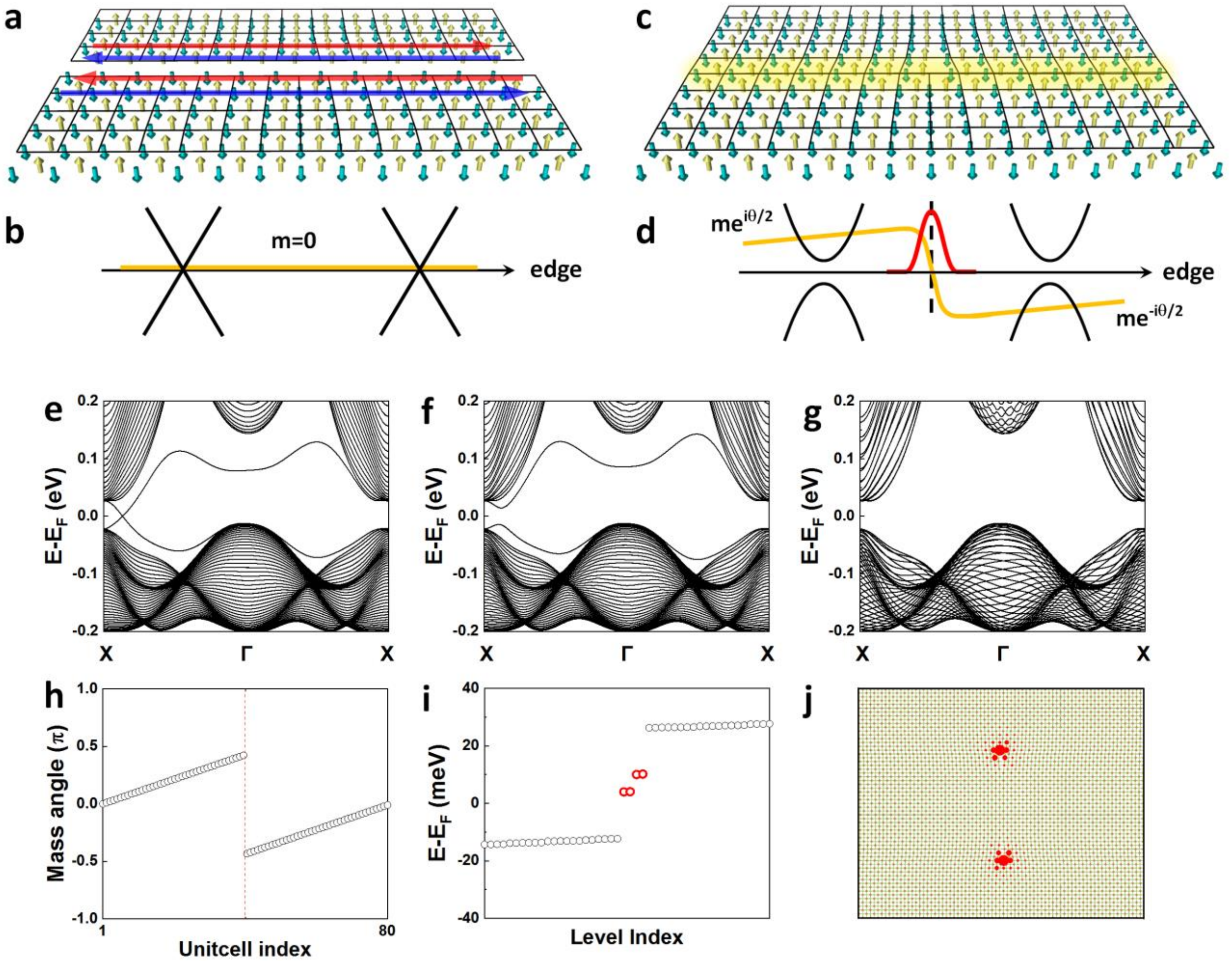

**Figure 5. Topological origin of the edge-dislocation state in FeSe. (a) Schematic of helical Dirac edge states along two domain edges without coupling. (b) Corresponding gapless Dirac edge states with zero mass ($m$ = 0). (c) Formation of an edge dislocation (yellow) by joining the top and bottom domains. (d) Mass-angle $m(\theta)$ shift across the dislocation (gold curve), generating an in-gap topological edge-dislocation state (red curve). (e–g) Band evolution of two FeSe nanoribbons with inter-edge hopping of 0, 30%, and 100%, respectively. (h) Mass-angle shift across the edge dislocation. (i) Discrete energy levels of the corresponding FeSe supercell, showing four in-gap topological states (red circles). (j) Real-space distribution of the dislocation states, localized at the dislocation cores.**

We assume the emergence of the local modes at edge dislocations can be treated by coupled edge states with different inter-cell hopping intensities. To validate this scenario, we performed tight-binding-based DFT calculations. Figures 5e-g and Figure S8 show the evolution of edge bands with different inter-cell hopping intensities for the two boundaries. As the inter-cell hopping intensity increases from zero (Figure 5e) to 20% (Figure 5f) and 100% (Figure 5g) of the normal value, the edge bands will completely merge into the bulk bands and form a gap feature, consistent with the experimental observation of the "gapped" feature in bulk FeSe films (Figure 3b).

We further find that even though the moduli of mass terms $m$ are invariant for the translational symmetry, their phases (mass angle $\theta$) change across the dislocation. As shown in Figure 5h, the numerical difference between mass angles from the two sides reaches its maximum when they meet at the center of the dislocation, but with different signs. As expected, the topological dislocation modes appear in the bulk band gap of FeSe, as depicted by red circles in Figure 5i and Figure S9. The calculated dislocation modes have a localization radius of approximately $4a_{Se}$, corresponding to 1.5 nm, which is close to the values obtained from Gaussian fits to the experimental dI/dV maps (2.2–2.7 nm), as shown in Figure S10. There are four modes in the system separated into two groups: each dislocation holds (Figure 5j) two modes with different spin polarizations and slightly split energies. This behavior is a consequence of time-reversal symmetry, so that these dislocation modes must appear as pairs with opposite spins propagating in opposite directions, thus resembling the chiral edge states of a 2D quantum spin Hall insulator. (The energy splitting is due to unequal perturbations of dislocations to states with different spin polarizations for the antiferromagnetic.) We also performed the results for 2 ML FeSe films, which are similar (Figure S11). Our DFT calculations find that edge dislocations potentially host half a fractional charge, which requires experimental and theoretical efforts for confirmation in the future.

Strain and charging effects are naturally present near a dislocation and are known to alter local electronic properties[46–48]. To evaluate the strain distribution around the edge dislocation, we performed the geometrical phase analysis (GPA)[46], shown in Figure S12. The GPA maps reveal tensile strains along the Se-Se lattice direction in the vicinity of the dislocation cores, with variations across sites likely due to local structural reconstructions (Figure 2f). However, the characteristic V-shaped $dI/dV$ spectra are observed at every edge dislocation, ruling out strain as the origin of these modes. While our calculations indicate that strain can shift the energies of the edge states relative to the Fermi level (Figure S13), it does not account for their robust spatial confinement. Charging effects at GBs[47,48], arising from defect or substrate-induced charge transfer, are known to shift band edges, as reported in $Bi_2Se_3$[41] or $MoS_2$[42]. However, such gap variations typically depend sensitively on the grain misorientation angle[42], which is not observed in this work. Having excluded strain and charging effects, we attribute the observed local states in FeSe(001)/$SrTiO_3$(111) films to topologically protected helical modes bound to the dislocation. Future spin-polarized STM or magnetic-field-dependent measurements would provide direct insight into the spin texture and field response of these dislocation states.

The symmetry-mismatched epitaxy strategy, which leverages rotational mismatch between film and substrate, is inherently material-agnostic and can be extended to a broad range of materials for controllable fabrication of topological defects, including topological materials[49], other high-temperature superconductors[50], and transition-metal dichalcogenides[51]. Looking forward, inducing superconductivity into multilayer FeSe films, for instance through alkali metal intercalation, provides a promising route toward realizing novel quantum phenomena, such as the emergence of Majorana zero modes[52], Josephson junction arrays[53–55], and even scalable topological qubit networks[56,57]. Notably, the naturally formed dislocation lines could serve as predefined channels for braiding operations, offering a potential pathway toward fault-tolerant topological quantum computing.

## CONCLUSIONS

We have successfully synthesized tetragonal FeSe(001) films on the $SrTiO_3$(111) substrates by MBE, where symmetry-mismatched epitaxy drives spiral growth and the formation of low-angle grain boundaries composed of periodic edge dislocations. Atomic-resolution STM imaging and spectroscopy reveal localized states at step edges and at the cores of both edge and screw dislocations. Calculations support a topological origin for these bound states, which may serve as conduits for dissipationless 1D transport[58] or as building blocks for topological quantum devices. In particular, FeSe films can be tuned into a superconductor by charge doping; thus, the interplay between superconductivity and defect-localized modes raises the intriguing possibility of realizing Majorana zero modes at the topological defects[52], a key goal in topological quantum computing. Beyond FeSe/$SrTiO_3$ system, our findings demonstrate that symmetry-mismatched epitaxy provides a tunable platform for investigating the rich physics that emerges at the intersection of real- and momentum-space topology.

## METHODS

***Thin-film growth.*** The FeSe films were grown by molecular beam epitaxial (MBE) on $SrTiO_3$(111) substrate at 300-400 °C. During the film growth, high-purity Fe (99.9999%) and Se (99.9999%) were evaporated from Knudsen cells heated to 1150 and 130 °C, respectively. Tshe epitaxy growth is performed under a base pressure of $5.0 \times 10^{-10}$ Torr.

***STM/S measurements.*** Low-temperature STM/S measurements were conducted on a Unisoku (USM1300) STM. After MBE growth, the FeSe(001)/$SrTiO_3$(111) films are transferred *in situ* to the STM for measurements at $T$ = 4.5 K with PtIr tips calibrated on Ag/Si(111) films. The dI/dV spectra were acquired using the standard lock-in technique by applying a bias modulation of 20 mV (r.m.s.) at 732 Hz, otherwise specified. All the STM images are collected under constant-current mode.

***First-principles density functional theory calculations.*** First-principles calculations, including spin-orbital coupling, were carried out in the framework of generalized gradient approximation with Perdew-Burke-Ernzerhof functional using the Vienna Ab initio simulation package (VASP). All calculations were performed with a plane-wave cutoff of 400 eV on 11 × 11 × 1 Monkhorst-Pack $k$-point mesh. The convergence criterion of energy was smaller than $10^{-5}$ eV. During the structural relaxation, all atoms were relaxed until the forces were smaller than 0.01 eV/Å. The vacuum layer was 15 Å thick to ensure decoupling between neighboring slabs. The first principles tight-binding Hamiltonian was obtained from band fitting by using the Wannier 90 package. The five d orbitals of Fe and three $p$ orbitals of Se were used in band fitting. The topological edge dislocation states were simulated by extending the unit cell Hamiltonian to a supercell configuration with only nearest-neighbor terms. An 81 × 30 periodic FeSe superlattice was built, containing a pair of edges to construct the dislocation using the cut-and-glue method. The discrete energy levels show four in-gap topological edge dislocation states. The real-space distribution of the edge-dislocation states localize around the dislocation cores with spatial radii of ~$4a_{Se}$.

## ASSOCIATED CONTENT

Supporting Information

The Supporting Information is available free of charge.

Note 1: Edge dislocation model and Bulk-dislocation correspondence; Figure S1. Comparison of the surface topography of FeSe films epitaxially grown on $SrTiO_3$(001) and $SrTiO_3$(111) substrates; Figure S2. STM topography of FeSe films epitaxially grown on the $SrTiO_3$(111) substrate; Figure S3. Low-angle grain boundaries with varying misorientation angles in FeSe/$SrTiO_3$(111); Figure S4. Layer-dependent dI/dV spectra of FeSe films epitaxially grown on $SrTiO_3$(111); Figure S5. Topographic STM image and corresponding dI/dV maps of edge dislocation arrays; Figure S6. Characterization of low-angle grain boundaries with varying misorientation angles in FeSe/$SrTiO_3$(111); Figure S7. Screw dislocations in FeSe/$SrTiO_3$ (111) films; Figure S8. Inter-edge coupling induced band gap for the bulk; Figure S9. Dislocation-induced localized modes in FeSe; Figure S10. Spatial distribution of dislocation state; Figure S11. Band structure of 2ML FeSe; Figure S12. Characterization of edge dislocation cores; Figure S13. Local stress-induced shift of dislocation state

AUTHOR INFORMATION

**Corresponding Author**

**Lian Li** − Department of Physics and Astronomy, West Virginia University, Morgantown, West Virginia 26506, United States

Phone: (+1) 304-293-4270; E-mail: lian.li@mail.wvu.edu

**Zhengfei Wang** − Hefei National Research Center for Physical Sciences at the Microscale, CAS Key Laboratory of Strongly-Coupled Quantum Matter Physics, Department of Physics, University of Science and Technology of China, Hefei, Anhui 230026, China

Phone: (+86) 0551-63607360; E-mail: zfwang15@ustc.edu.cn

## Authors

**Huimin Zhang** − Department of Physics and Astronomy, West Virginia University, Morgantown, West Virginia 26506, United States

Key Laboratory of Materials Modification by Laser, Ion and Electron Beams (Dalian University of Technology), Ministry of Education, Dalian 116024, China

**Basu Oli** − Department of Physics and Astronomy, West Virginia University, Morgantown, West Virginia 26506, United States

**Michael Weinert** − Department of Physics, University of Wisconsin, Milwaukee, WI 53201, USA

## Author Contributions

L.L. and H.Z. conceived and organized the study. H.Z. and B.O. performed the MBE growth and STM/S measurements. Z.W. carried out the DFT calculations. All authors analyzed the data, and H.Z. and L.L. wrote the paper.

## Notes

The authors declare no competing financial interest.

## ACKNOWLEDGMENTS

HZ acknowledges support from the National Key Research and Development Program of China (2023YFB3809600), National Natural Science Foundation of China (Grant Nos. T2495211, 12304210), Young Top Talents in "Xingliao Talent Program" of Liaoning Province (XLYC2403120), Natural Science Foundation of Liaoning Province (2026-MS-014), Open Research Fund Program of the State Key Laboratory of Low-Dimensional Quantum Physics, and Xiaomi Young Talents Program. ZW acknowledges support from the National Natural Science Foundation of China (Grant Nos. 12574526, 12174369), and thanks the Supercomputing Center at USTC and Hefei Advanced Computing Center for providing computing resources. MW and LL acknowledge support from the U.S. Department of Energy, Office of Basic Energy Sciences, Division of Materials Sciences and Engineering under Award No. DE-SC0017632s.